\documentclass[printer]{tMOP2e}

\begin{document}
\doi{10.1080/0950034YYxxxxxxxx}

\markboth{Williams \& Saffman}{Compression and localization of an atomic cloud in a time dependent optical lattice}

\title{Compression and localization of an atomic cloud in a time dependent optical lattice}

\author{Will Williams$^\dagger$ \thanks{$\dagger$ E-mail:wdwilliams@wisc.edu} \& M. Saffman\\
Department of Physics, 1150 University Avenue, University of Wisconsin, Madison, WI, 53706}

\maketitle

\begin{abstract}
We analyze a method of compressing a cloud of cold atoms  by dynamic control of a far off resonance optical lattice.  We show that by reducing the lattice spacing either continuously or in discrete steps while cooling the atoms with optical molasses large compression factors can be achieved.  Particle motion in the time dependent lattice is studied numerically using a three dimensional semiclassical model. Two experimentally realistic models are analyzed. In the first we continuously vary the lattice beam angles to compress atoms initially in a Gaussian distributed cloud with standard deviation of $250~ \mu\rm m$ into a single site of a two-dimensional lattice of area $A\sim 35\times 35 ~\lambda^2$, with $\lambda$ the wavelength of the lattice  beams.  This results in an optical depth for an on-resonant probe beam $>80$  which is an increase by a factor of about 1800  compared to the uncompressed cloud.  In the second approach we use a discrete set of lattice beam angles to decrease the spatial scale of the cloud by a factor of 500, and  localize a few atoms to a  single lattice site with an area $A\stackrel{<}{\sim}~\lambda^2.$

\end{abstract}

\section{Introduction}
\label{Intro}

During the last few decades there has been intense development of methods to cool and localize atomic samples with optical fields. 
Conservative trapping potentials can be created using far off resonance light beams that are focused to a small spot\cite{Chu1986,Heinzen1993}. These far off resonance traps  have become widely used
as a starting point for achieving Bose Einstein 
condensation\cite{Chapma2001,Kinosh2005,Raizen2005oe}, as well as for experiments with single atoms\cite{Grangi2004,Mesche2004,Yavuz2005}. There is currently much interest in techniques for loading a large number of atoms at high densities into optical traps\cite{Walker2005} and in methods of controlling the loading of an optical lattice\cite{porto}. These capabilities are important for creating large condensates, as well as highly localized atomic ensembles for quantum logic and 
communications\cite{Saffma2005b}.   The phase space density of atomic clouds can be increased using
laser or evaporative cooling as well as by novel approaches such as  asymmetrical optical barriers\cite{Raizen2005}. Heating due to photon rescattering limits the phase space density that can be reached by optical cooling  although spatially anisotropic samples\cite{Vengal2003} 
as well as optimized cooling techniques in optical 
lattices\cite{Winoto1999,Wolf2000} have led to improved performance.
Dynamically changing the optical potential in conjunction with optical cooling can also increase phase space density as in experiments which compressed an atomic cloud  using  a trap created by rapidly rotating  a single blue detuned beam\cite{Rudy2001,Friedm2000}.

In this paper we use numerical simulations to study the performance of a novel  protocol for transferring a three-dimensional distribution of atoms into a single well of a planar lattice. The anisotropic  well is  elongated perpendicular to the plane of the lattice which minimizes heating and repulsion due to photon rescattering\cite{Vengal2003} and is well suited for applications that require mode matching between a Gaussian laser beam and a pencil shaped atomic ensemble\cite{Saffma2005b}. The technique can also be extended to a three dimensional lattice to create a spherically shaped localized cloud of atoms.  

The proposed method starts with a large period optical lattice which 
confines all atoms in a single trapping site. The angles between the beams are then increased to create wavelength  scale lattice sites. We show that if the lattice beam angles are changed slowly, and the atoms are simultaneously cooled by optical molasses, it is possible to keep a large fraction of the atoms in a single highly localized  trapping site. 
  Atoms of temperature $T$ are localized within a distance  $\delta r\sim w\sqrt{k_B T/U}$ from the origin in a well of depth $U$ that is created by a focused laser beam with  Gaussian waist (radius to the $1/e^2$ intensity point) $w$.    The atoms have a vibrational oscillation frequency of $\nu=(2/w)\sqrt{U/m},$ where $m$ is the  mass of an atom.
 Provided the well size is reduced on a time scale that is long compared to the oscillation frequency the atoms will tend to be trapped in the center of the well and get compressed.  Under this adiabatic condition optical molasses is not necessary to spatially compress the atoms but there will be no increase in phase space density. By combining  dynamic lattice reshaping with optical cooling the atoms can be compressed to higher phase space density. 

\begin{figure}[!t]
\begin{center}
\includegraphics[width=15cm]{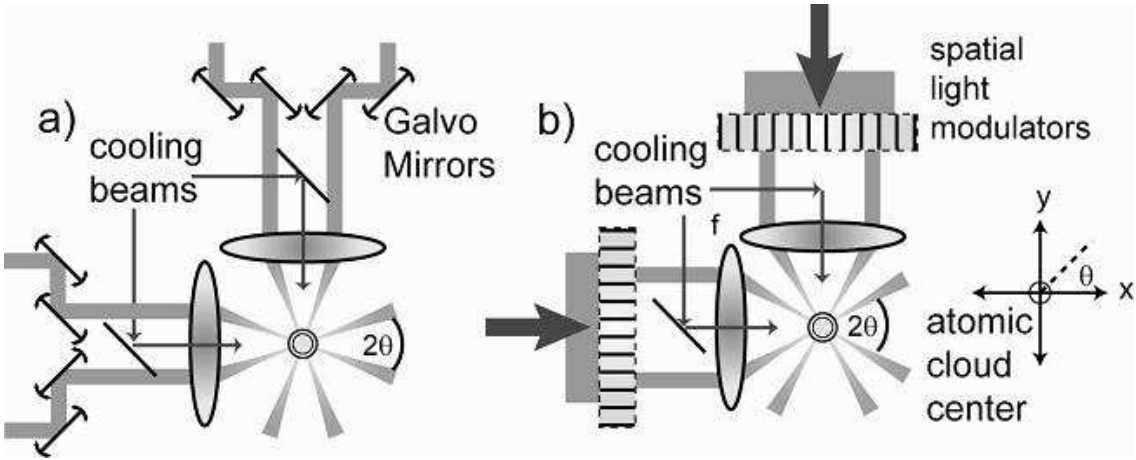}
\caption{\label{fig:schematic}Arrangements for dynamic compression by varying the angles between lattice beams: a) The beam crossing angles are changed continuously using four  pairs of galvanometer mounted mirrors, b) The beam crossing angles are changed in discrete steps using two  SLMs.  The additional cooling beams provide optical molasses during the compression.}
\end{center}
\end{figure}

In this paper we numerically explore two methods for manipulating atomic samples using dynamic optical lattices.  The first method assumes a continuously variable angle between lattice beams, as in  Fig \ref{fig:schematic}a, to compress a large cloud in two dimensions  to provide a large optical depth along the third dimension. Continuously variable angles could be obtained using, for example, galvanometer mounted mirrors or acousto-optic modulators. 
  The second method uses a discrete set of beam angles, as could be implemented with one-dimensional spatial light modulators (SLM's) with $N$ pixels, see Fig \ref{fig:schematic}b. 
 In the first approach we use relatively small compression factors of about 15 to create pencil shaped 
 atom clouds with a large optical depth. In the second approach we can achieve a compression factor of about 500 and localize atoms to a single lattice site with area $A\stackrel{<}{\sim} ~\lambda^2$, with $\lambda$ the wavelength of the lattice beams.

The rest of the paper is organized as follows. 
The  theoretical and numerical model for the  dynamical optical lattice are detailed in Sec. \ref{opttheory}. Section \ref{numerics} presents results of numerical simulations that validate the proposed compression technique for continuous variation of the  beam angles, as well as discrete steps. 
A comparison between traveling wave and standing wave lattice geometries is also given.
We conclude with Sec. \ref{discussion}  which gives a discussion of  applications and limitations of the technique.

\section{Theory}
\label{opttheory}

\indent We describe a method to compress atoms in a single site of a two-dimensional lattice by dynamically changing the angles between the lattice beams.  The conservative optical potential for a two-level atom including the effects of saturation is\cite{McClel1995}
\begin{equation}
\label{eqpotential}
U({\bf r})=\frac{\hbar \Delta}{2}\ln\left[1+ \frac{I({\bf r})}{I_s}\frac{1}{(1+4\Delta^2/\gamma^2)}\right],
\end{equation}
where $I({\bf r})$ is the optical intensity  at position ${\bf r}$, $I_s$ is the saturation intensity, $\gamma$ is the natural linewidth, $\Delta=\omega-\omega_a$ is the detuning from resonance, $\omega$ is the optical frequency, and $\omega_a$ is the atomic transition frequency.

Consider two equal intensity plane traveling waves with wavevectors ${\bf k}_j=k(\cos\theta_j \hat x + \sin\theta_j \hat y)$ where $k=2\pi/\lambda,$  $\lambda$ is the wavelength,  
and $\theta_j$ is the angle each beam makes with the $x-$axis. 
When $\theta_1$=-$\theta_2$=$\theta$ the angle between the beams is $2\theta$ and the intensity is  
\begin{equation}
\label{intensitystep1}
I(x,y) = 4 I_0 \cos^2\left[\pi y/\Lambda(\theta) \right]
\end{equation}
where $I_0$ is the intensity of each wave and the spatial  period is  $\Lambda(\theta)=\frac{\lambda}{2 \sin(\theta)}$.  To achieve compression in one dimension $\theta$ is increased from 0 to $\pi$/2 radians.  Doing so  decreases the periodicity from infinity (a plane wave with uniform intensity) to $\frac{\lambda}{2}$, the periodicity of a single standing wave.  To accomplish two dimensional compression another set of traveling waves are added with their bisector orthogonal to the bisector of the first pair as in the configurations shown in Fig. \ref{fig:schematic}.  By arranging for the second pair of waves to be polarized orthogonally to the first pair we 
obtain\footnote{Superposition of the orthogonally polarized pairs of beams creates a field that has a spatially varying polarization state. In a real multilevel atom the optically induced potential depends on the polarization when there is a nonzero vector or tensor polarizability. We assume in this paper that polarization effects are negligible so that the potential is accurately described using the intensity of Eq. (\ref{intensitytravel}). This is  true for alkali atom ground states provided the wavelength is sufficiently far detuned from the strong D lines. } 
\begin{equation}
\label{intensitytravel}
I(x,y) =4I_0 \left\{\cos^2\left[\pi x/\Lambda(\theta)\right]+\cos^2\left[\pi y/\Lambda(\theta)\right]\right\}.
\end{equation}

\begin{figure}[!t]
\begin{center}
\includegraphics[width=12cm]{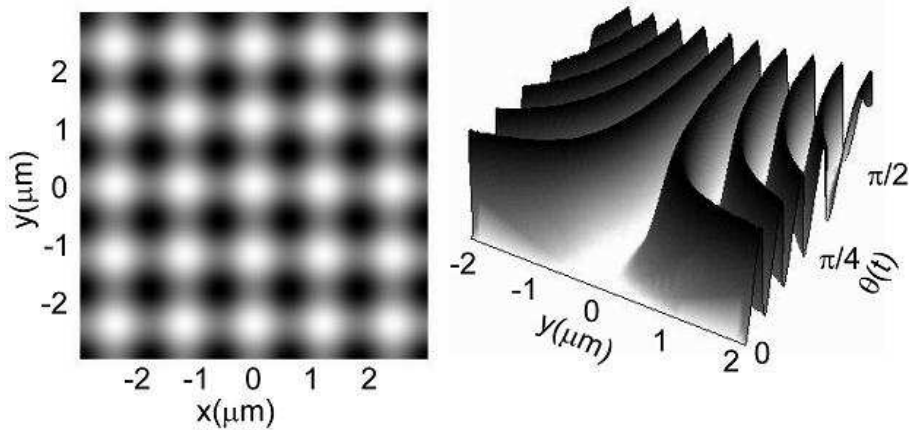}
\caption{\label{fig:area}Left: Density plot of the interference pattern created by four traveling waves, as described in the text, with $\theta=\pi/8$. Regions of higher intensity are shown in white. 
 Right: The intensity along the $y-$axis as a function of $\theta$. When $\theta=0$, there is a plane traveling wave with equal intensity over all space.   As $\theta$ increases, a periodic structure forms and the well area decreases.   }
\end{center}
\end{figure}

This intensity produces a two dimensional lattice pattern with periodicity of $\Lambda$ in both the $x$ and $y$ directions.  Each site of the lattice has an area of $A(\theta)\approx \Lambda^2=\frac{\lambda^2}{4 \sin^2(\theta)}$.  As $\theta$ increases from $0$, the area of each well decreases as shown in  Fig. \ref{fig:area}. 
If the initial value of $\theta$ is small enough the central well will be large enough to contain essentially all of the atoms. 
Any atoms that start in the central  well will stay there during the compression of the wells as long as they are not energetic enough to travel over the intensity barrier into the next well.  If some of the atoms escape the central well, the method can be repeated to obtain better compression. A similar protocol is also possible using 
standing waves as will be discussed in Section \ref{numerics}.

The compression protocol thus involves the following sequence. We start with a cloud of cold atoms and superimpose a two-dimensional lattice of optical trapping beams created by two pairs of crossing beams propagating along $\hat x$ and $\hat y$ as shown in Fig. \ref{fig:schematic}. The initial value of $\theta=\theta_{\rm min}$ is chosen small enough such that almost all the atoms are confined in the central lattice well. We then 
increase $\theta$, either continuously or in discrete steps, which spatially compresses the atoms but also heats them since the well area decreases. The increase in temperature is removed by applying optical molasses during the compression, either continuously, or for a time $T_c$ between each step.  The compression sequence is applied  until the atoms have been localized into a lattice site with small area created by the maximum crossing angle $\theta=\theta_{\rm max}.$

\section{Numerical Results}
\label{numerics}

Atomic compression was simulated numerically using Newtonian dynamics. The equation of motion $m\ddot{{\bf r}}=-\nabla U$ was used to update the velocity and position of each atom during the compression protocol. 
Atomic collisions were not accounted for.
Cooling by three dimensional molasses was simulated by testing the probability that a photon is scattered at every time step using an approximation to the three dimensional distribution of the optical field   as described in \cite{Wohlle2001}.  If the atom scattered a photon,  the atom's momentum was changed by $\hbar$k along the direction of the appropriate cooling beam for absorption and then in a random direction in three dimensions for emission. The dipole emission pattern of a real atom was approximated by a spherically uniform emission probability.    The atomic parameters were chosen to correspond to the 
D2 transition of Cs ($6^2S_{1/2}-6^2P_{3/2}$) at  $\lambda=852~\rm nm$,  with excited state decay rate $\gamma=2\pi\,\times 5.22~\rm MHz$ (this was reduced by a factor of 3 as explained below). The intensity of each cooling beam was set to  $I/I_s=0.1~(1.09~\rm W/m^2)$  and the detuning was set to  $\Delta=-\gamma/2$.  To avoid the effects of differential ground and excited state AC Stark shifts due to the lattice beams, the cooling  and lattice beams were flashed on and off  out of phase with each other at a frequency of $4~\rm  MHz$.  The rate was  sufficiently large compared to the highest vibrational frequency encountered  to avoid loss of atoms during the periods when the lattice beams were off.  SubDoppler cooling was simulated by reducing the natural linewidth by a factor of three.  The Doppler temperature for 6 cooling beams each with $I/I_s=0.1$ and a $\times 3$ reduced linewidth is about $54 ~\mu \rm K$ in the $x,y,$ and $z$ directions.

  In the following sections we give numerical examples of atomic compression with cooling for continuously variable beam angles (Sec. \ref{sec.galvo}) and discrete angles (Sec. \ref{sec.slm}).  We have used geometrical parameters that are compatible with future experimental demonstrations.  
During the compression the maximum atomic density is limited partly by collisions (which are not accounted for  here) and  primarily by  radiation pressure from  on-resonance spontaneously emitted photons.  This limits the density in a magneto-optical trap (MOT) to a few times $10^{11} \rm{atoms/cm^3}$.  However, in the lattice, the atoms are in a cylindrical pencil shaped trap that gets more elongated as the compression proceeds.  Due to this pseudo one dimensional geometry the limitations due to radiation pressure are reduced. We discuss the maximum compression achievable given radiation pressure effects in  Sec. \ref{discussion}.

\subsection{Continuous variation of beam angles}
\label{sec.galvo}

In this section we assume continuous variation of the beam angles using the galvanometer mirror implementation of Fig. \ref{fig:schematic}a .   The mirrors are placed such that the beam separation varies from $1~\rm mm$ to $15~\rm mm$ in front of a $f=50~\rm cm$ focal length lens that focuses the beams in to the atomic cloud.   The beam waists in the front focal plane are set at $250~ \rm{\mu m}$ in the $x-y$ plane,  and $75~ \rm{\mu m}$ along $\hat z$ (axial direction) giving  $w_{\rm x-y}=0.54~\rm{mm}$ and $w_{\rm z}=~1.8~\rm{mm}$ at the atoms.  Thus in the galvanometer mirror setup, the half beam crossing angle can be changed from $0.057 \le \theta \le 0.86~\rm deg., $ giving a lattice periodicity of 
$430~\rm{\mu m} \ge \Lambda \ge 28 ~\rm{\mu m}$.  This is an overall compression ratio of about 15.

The lattice beam detuning was set to $\Delta=-2 \pi\,\times 15~\rm{GHz}$ and the peak intensity was $2 \times 10^6 ~\rm{W/m^2}$ which results in a well depth of $U_0/k_B=2~\rm  mK$ with an effective well depth from peak to the interwell ridge of $1~\rm mK$.  This choice of detuning and intensity  gives a scattering rate of $9.1 \times 10^4 ~\rm{photons/sec}$ and requires a total power of $1.54~\rm W$ for each pair of beams.  This high power is a result of using both a deep well and our choice of beam waists.  The transverse beam waist was chosen such that the  atomic cloud would fit into a single site of the initial large period lattice.  The axial beam waist was chosen so that the compressed cloud could be elongated along $\hat z$ giving a large optical depth.  Relaxing any of these parameters will result in lower power requirements.  The cooling beams give a scattering rate of $3.8 \times 10^6 ~\rm{photons/sec}$ which is a factor of about 42 larger than the scattering rate due to the optical lattice, which therefore only perturbs the cooling weakly.

\begin{figure}[!t]
\begin{center}
\includegraphics[width=13cm]{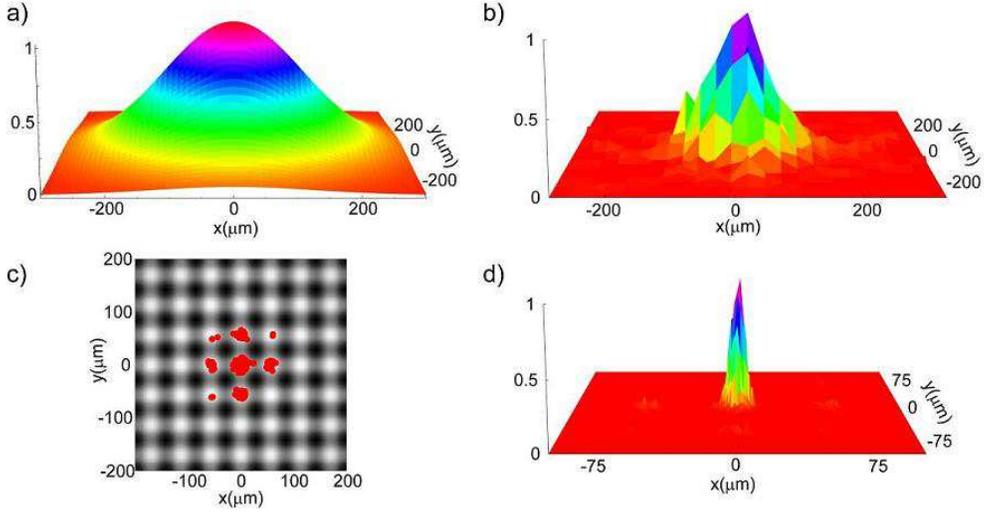}
\caption{\label{fig:final2}(Color Online) Compression with continuous change of beam angles:  a)The initial Gaussian density distribution, b) the density after being placed in the trap and cooled for $4~\rm ms$,  c) a contour plot showing the final atom distribution in the lattice, and d) the final number density.
}
\end{center}
\end{figure}

\begin{figure}[!t]
\begin{center}
\includegraphics[width=15cm]{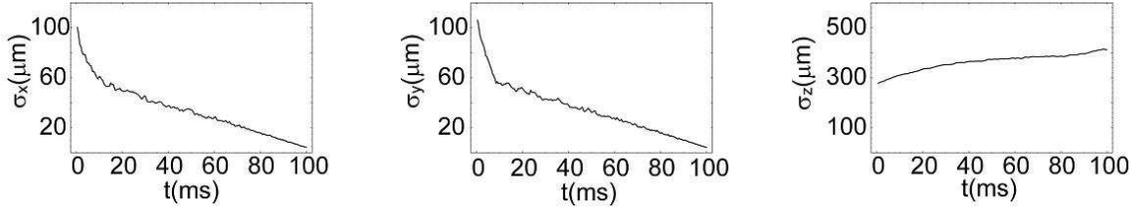}
\caption{\label{fig:final3} The standard deviations of the atomic cloud along all three axes during the compression.
}
\end{center}
\end{figure}

The initial atom distribution was a Gaussian with standard deviations of $\sigma_x=\sigma_y=\sigma_z=250~ \mu\rm m.$  The initial temperature was set to  $T_a=140~ \mu\rm K$ in $x, y$ and $z$.  The cloud was then placed in the potential defined by the first step in the lattice compression and cooled for $4~ \rm{ms}$.  The cloud was then compressed by continuously changing $\theta$ between the minimum and maximum values given above over 
 a time of $100 ~\rm{ms}.$ This produced a minimum lattice site area of  $A(\rm \theta_{\rm max}) \simeq  810 ~\mu \rm m^2$.
The results of a simulation  run with 1000 atoms are shown in Fig. \ref{fig:final2} and the change of the cloud widths with time are shown in Fig. \ref{fig:final3}.    The final compression captured 89\% of the atoms in the central well with the rest of the atoms  in the neighboring wells.  The loss of atoms in the central well is mainly due to the initial placement of the Gaussian cloud in the optical potential.

\subsection{Discrete set of beam angles}
\label{sec.slm}

In this section we vary the beam angles in discrete steps as could be produced by 
a pixelated SLM followed by a lens for focusing  into the region containing the atomic sample as shown in Fig. \ref{fig:schematic}b.
The SLM with rectangular pixels is  illuminated with an elliptically shaped Gaussian beam that is thinner than the height of each pixel along $\hat z$ but covers all of the SLM pixels in the $x-y$ plane.
The lenses between the SLM and the atom trapping region provide a Fourier transform of the field distribution so that switching  a single SLM pixel into the transmitting state results in a beam in the trapping region with a Gaussian envelope along $z$ (perpendicular to the plane of the lattice), and a sinc squared envelope in the $x-y$ plane. Provided that the width of the sinc function is at least a few times larger than the size of the lattice site created at $\theta_{\rm min}$ the numerical results  depend only weakly  on the exact form of the transverse field distribution. 
Comparing the  difference between the intensity pattern created by a plane wave and a plane wave enveloped by a sinc squared function whose electric field has a ${\rm FWHM}=200~\mu\rm m$, the only noteworthy difference is on the first step.  For the first step, the atoms are not confined near the bottom of the well and the FWHM of the  sinc squared  is smaller than the FWHM of the  central well created by plane waves by about $48\%$ 
which  results in a larger number of atoms lost on  the first step.  
After step 1, the percent difference of the FWHM's of the plane-wave and sinc squared models, drops rapidly and  is below $1\%$ by step 5.  Since the atoms are confined near the bottom of the well after step 1, the small difference between the two potentials is negligible. We have therefore used interfering plane waves in the $x-y$ plane for the numerical work, and assumed the focused beams were circular with waists of $w_z=w_y=100~\mu\rm m$  (for e.g. a beam propagating along $\hat x$) in  calculations of optical power requirements.

The beam detuning was set to $\Delta=-2 \pi\,\times 5~\rm{GHz}$ and the peak intensity was $2 \times 10^6 ~\rm{W/m^2}$ which results in a well depth of $U_0/k_B=6~\rm  mK$ with an effective well depth of $3~\rm mK$.  This choice of detuning and intensity  gives a scattering rate of $7.8 \times 10^5 ~\rm{photons/sec}$, and requires a total power of $5~\rm W$ for each pair of beams. 
 This high power is a result of illuminating the entire SLM with a cylindrical beam whose parameters fit the width of the SLM, but only opening two pixels at a time. The scattering rate from the lattice beams is about $20~\%$ of the scattering rate of the cooling beams. This ratio could be reduced further by using higher powers and larger detunings.  

\begin{figure}[!t]
\begin{center}
\includegraphics[width=10cm]{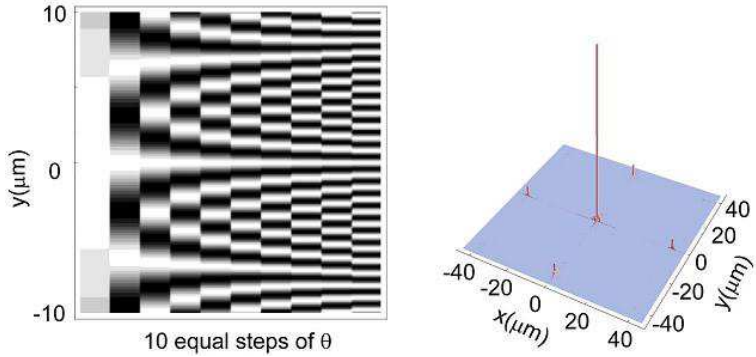}
\caption{\label{fig:pullaway}(Color Online) Left: Density plot of the intensity of the trap along the $y-$axis as $\theta$ changes by equal increments.  Right: Atomic density at the end of a compression sequence with $\theta$ changing by 0.7 deg. at each step. If an atom is at  a local potential minimum that is not the central well, it will end up far from the central well, as shown by the secondary peaks.}
\end{center}
\end{figure}

\begin{figure}[!t]
\begin{center}
\includegraphics[width=10cm]{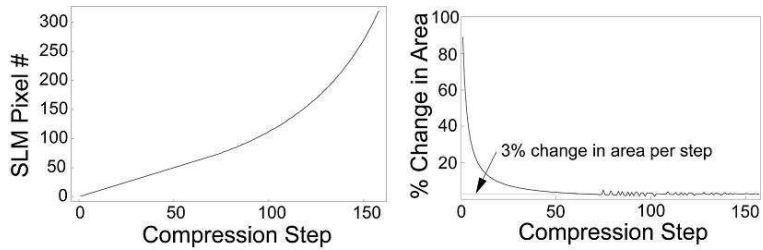}
\caption{\label{fig:anglechange}Left: The number of the SLM pixel turned on at each step in the compression sequence.   This represents the best possible fit allowed by a 640 pixel SLM for a 3\% change in well area at each step.  Right: Percent change in well area for the compression scheme as a function of compression step.}
\end{center}
\end{figure}

When  using a discrete set of angles  there is a finite resolution with which the angle $\theta$ can be set. If $\theta$ is abruptly changed by too large an amount atoms may get trapped in a local minimum of the optical potential, and not be swept towards the center of the lattice as illustrated 
in  Fig. \ref{fig:pullaway}.  
In addition if the initial value of $\theta$ is not small enough some of the atoms will not start in the central well and will not end up at the origin. To avoid these problems a large number of SLM pixels are needed so that the steps in $\theta$ are small.  It is convenient if the amount of heating incurred at each step is approximately constant, since  $T_c$ can then be held constant. This will be the case if the fractional change in the area of each lattice site is the same for each step.  In an experiment  with a finite number of SLM pixels this requirement cannot be met exactly. For the simulations shown below we assumed the SLMs shown in Fig. \ref{fig:schematic} each had 640 pixels and that the range in $\theta$ was from $\theta_{\rm min}=0.063$ to 
$\theta_{\rm max}=35$ deg. The pixels used at each compression step were chosen to approximate as well as possible a 3\% change in well area between steps, as shown in Fig. \ref{fig:anglechange}.  This corresponds to using only 158 pixels on the SLM or 157 steps of compression.

Optical cooling is again necessary since the atoms heat as they are being compressed.  The cooling parameters are  the same as in the galvanometer mirror setup except we cool to the Doppler temperature.  The temperature of the atoms was monitored and when either the  cooling time exceeded  3 ms or all three Cartesian components of the temperature were below 
$180 ~\mu\rm K,$ the next step in compression was taken.  The Doppler temperature for 6 cooling beams each with $I/I_s=0.1$ is about $160 ~\mu \rm K$ in the $x,y,$ and $z$ directions.

\begin{figure}[!t]
\begin{center}
\includegraphics[width=13cm]{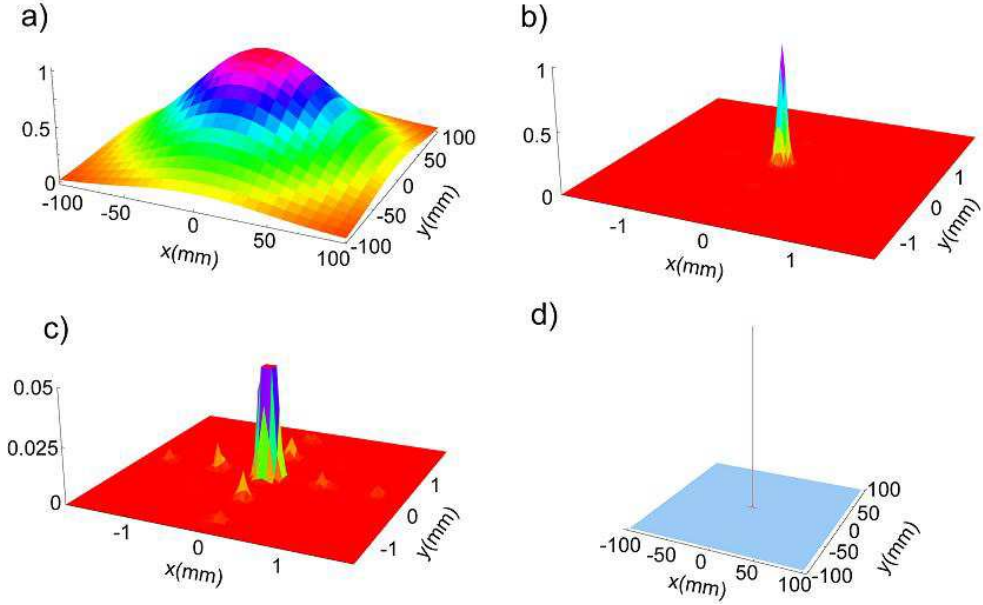}
\caption{\label{fig:final}(Color Online) Compression with discrete variation of beam angles: a) The initial Gaussian distribution, b) the final number density in a region of $ 25\times 25$  lattice sites , c) the final number density  plotted to a maximum of 5\% to show the atoms that have escaped into adjacent wells,  and 
d)  a histogram with bins defined by the lattice area $A(\rm \theta_{max})$ to display the final compression results.
}
\end{center}
\end{figure}

The initial atom distribution was a Gaussian with standard deviations of $\sigma_x=\sigma_y=\sigma_z=50~ \mu\rm m.$  The initial temperature was set to  $T_a=140~ \mu\rm K$ in $x, y$ and $z$.  The cloud was then placed in the potential defined by the first step in the lattice compression.    The simulation involves 157 steps 
and $\theta$ was varied between 0.063 and 35 deg. (a maximum of  70 deg. between two beams). These values correspond to using a Jen Optik SLM-S 640/12  with 640 pixels followed by a $f=45.6~\rm mm$ focusing lens.  We chose $\theta_{\rm max}=35$  deg. to correspond to  an experiment where there may only be 70 deg. of access on each side of the trapping region.  This will produce a minimum lattice site area of  $A(\rm \theta_{\rm max}) \simeq  0.55 ~\mu \rm m^2$.

\begin{figure}[!t]
\begin{center}
\includegraphics[width=12cm]{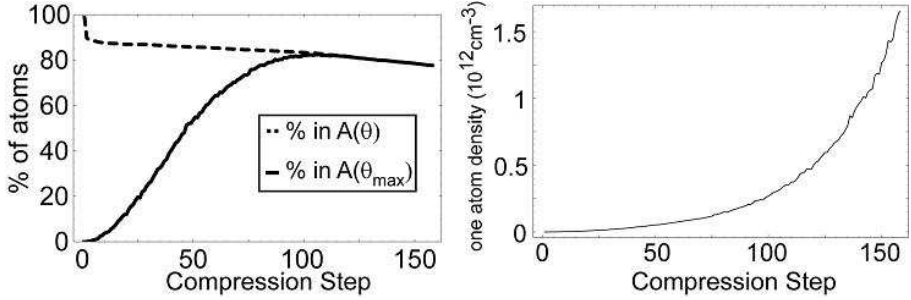}
\caption{\label{fig:captured}Left: Percent of atoms captured within the minimum 
well area $\rm{A(\theta_{max})}$ (solid line) and   percentage of atoms captured in the central well(dashed line). Right:  density per atom during the compression.}
\end{center}
\end{figure}

The results of a simulation  run with 5000 atoms are shown in Fig. \ref{fig:final}.  The cooling and compression time per step was varied from a maximum  of $T_c\simeq 3~\rm ms$ at the start of the compression sequence, to a minimum of $100~\mu\rm s$ at an intermediate stage, and was  $\sim 1.2~\rm ms$ per step near the end of the sequence, giving a  total  cooling and compression time of $115~\rm ms$.  The cooling time per step  at the beginning of the compression scheme is longer since the change in well area is larger.
The final compression captured 77.7\% of the atoms in the central well with 84.8\% of the atoms within a 20 $\mu$m square around the central spot. 
The loss of atoms in the central well is due to the initial placement of the Gaussian cloud in the optical potential, the temperature of the atoms, and the discrete nature of the SLM.  The largest loss of atoms occurs in the first stage of compression where the percent change in well area is about 88.8\%.  About 9\% of the atoms are lost in the first step, as seen in  Fig. \ref{fig:captured}. This initial loss could be reduced at the expense of higher laser power by having more SLM pixels or deeper wells. Once the cloud is safely contained in the lattice site, there is a steady decrease in atoms since the effective well depth of the trap is 3 mK and the temperature of the cloud is kept around $175 ~\mu\rm K$, which leads to a nonzero escape rate. 
  Even though the total number captured in the central well decreases, the density of the atoms, defined as the number of atoms within the central well divided by the volume of the well, still increases after every step as shown in  Fig \ref{fig:captured}.

\subsection{Standing wave lattices}
While the traveling wave lattices described above have a relatively simple experimental implementation, the intensity pattern defined by Eq. (\ref{intensitytravel}) only falls to 50\% between wells.  This effectively wastes half of the well depth requiring better cooling and more powerful lasers.
An alternative approach relies on retro-reflecting the beams to create standing waves.  Keeping the same geometry as the two traveling waves case, but including retroreflections,  the result of interfering all four beams comprising the two  standing waves is the intensity pattern
\begin{equation}
\label{intensitystep2}
I(x,y) \sim \cos^2\left[k x \cos(\theta)\right]\cos^2\left[k y \sin(\theta)\right].
\end{equation}

\begin{figure}[!t]
\begin{center}
\includegraphics[width=13cm]{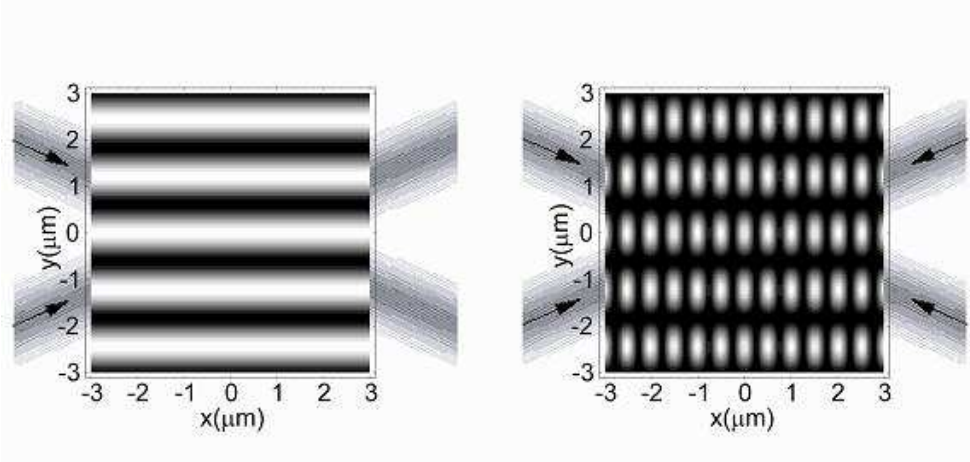}
\caption{\label{fig:compare}Comparison of two interfering traveling waves (left) and two interfering standing waves (right) with high intensity shown in white.  Both interference patterns are created with $\theta=\pi/8$ giving $2\theta=\pi/4$ between beams.   The traveling wave produces a 1D interference pattern while the standing wave produces a 2D lattice structure.  The incident  beam propagation directions are shown superimposed behind the plots.}
\end{center}
\end{figure}

\begin{figure}[!t]
\begin{center}
\includegraphics[width=10cm]{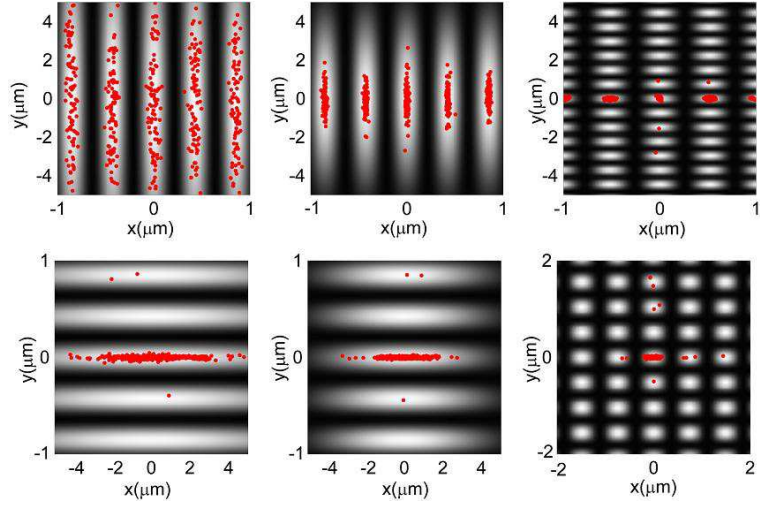}
\caption{\label{fig:stand}(Color Online) Compression sequence  using standing waves.  The first step (top row, left to right) uses two pairs of standing waves with bisector along $\hat x$.  The second step (bottom row, left to right) uses two pairs of standing waves with bisector along $\hat y$. This simulation uses the same optical parameters as in Section \ref{sec.slm}.}
\end{center}
\end{figure}

As seen in Fig. \ref{fig:compare} this results in a two-dimensional intensity pattern, as opposed to two-traveling waves which give a one-dimensional lattice.   Two-dimensional compression can be achieved by compressing sequentially along different axes as shown in  Fig. \ref{fig:stand}. In the first step, a pair of beams with bisector along $\hat x$ is separated  which  compresses the atoms along the $\hat y$ axis, while they fill   wells which are distributed along $\hat x.$  Then the process is repeated for a pair of beams with bisector along $\hat y$.   Since, after the first step, all of the atoms are located in wells along the $x$-axis, at the beginning of the second step, all of the atoms should be trapped in the central trough of the 1D standing wave on the $x$-axis.  As $\theta$ increases, the area of the wells decreases and all the atoms will be compressed to a single well located at $x=y=0$ with well area of $A=\lambda^2/2$ for $\theta_{\rm max}=\frac{\pi}{4}$.  In contrast to  the traveling wave case  the wells are  separated by regions of zero intensity which is a more efficient usage of laser power.   Experimentally, however, a retroreflection arrangement is more difficult to implement.

\section{Discussion}
\label{discussion}

The final density of the compressed atoms is limited by both photon rescattering and atomic collisions.  Typical high density MOTs provide densities $n_{\rm a}\sim 10^{11} ~\rm{cm^{-3}}$ while dark MOTs\cite{Ketter1993} and dark optical lattices\cite{Elman2005} have been used to reach  $n_{\rm a}\sim10^{12}~ \rm{cm^{-3}}$. Densities as high as  $n_{\rm a}\sim 10^{12} ~\rm{cm^{-3}}$ have also been obtained in a  highly anisotropic elongated bright MOT\cite{Prenti2004}. 
 We therefore expect that the final density achievable by the galvanometer mirror scheme is on the order 
of $n_{\rm a}=10^{12}~ \rm{cm^{-3}}$.  This is a conservative estimate for the SLM scheme because the final trap is very anisotropic which results in greatly reduced radiation pressure.  Using this conservative limit, it is only possible to fit a few atoms into the central well.
 If the MOT initially has too many atoms they will spread to other wells during compression resulting in many filled wells instead of one filled central well. The SLM compression scheme may therefore be suitable for loading a small number of  atoms into sites of an optical lattice with sub wavelength periodicity. In this context the compression scheme  provides an effective zoom lens with a magnification ratio of about 500  which would be difficult to achieve with a conventional design.   Additional numerical simulations show that 
it is possible to direct the atoms into a preferred lattice site, instead of the central site, by introducing a small phase shift between the beams at each compression step.

The galvanometer mirror setup with a relatively small $\theta_{\rm max}$ is useful for creating atomic ensembles with large optical depth.
This is of interest for studying quantum state mapping between light and small atomic ensembles\cite{Duan2000}. The initial spherically symmetric MOT cloud has an optical depth, for a narrow beam with a width that is small compared to the MOT,  of 
$d = \int_{-\infty}^\infty dz\, \sigma n_a(x=0,y=0,z)$ where $n_a(x,y,z)$ is the atomic density and $\sigma=3\lambda_a^2/(2\pi)$ is the resonant cross section for a two-level atom with transition wavelength $\lambda_a.$
Assuming  a MOT with Gaussian density distribution and $\sigma_x=\sigma_y=\sigma_z=250~\mu\rm m$ as in the previous section, and $N=5 \times 10^4$ atoms we find $d=0.044,$ which would imply a very weak interaction between a probe beam and the MOT atoms.

  However, $80\%$ of these same $5 \times 10^4$ atoms can be rapidly compressed into a pencil shaped lattice site to achieve an optical depth of 
$d>80$, which would provide a strong interaction between the probe light and the atomic sample.  To show this we examine the regime where the atomic temperature is small compared to the well  depth as in our numerical simulations. The atomic density distribution inside each elongated well will then be Gaussian with 
$\sigma_x=\sigma_y=[\Lambda(\theta)/\pi]\sqrt{k_B T_a/U_0},$ and $\sigma_z=(w_z/2)\sqrt{k_B T_a/U_0}.$
The atomic density distribution can then be approximated by 
\begin{equation}
n_a(x,y,z) = \frac{N}{(2\pi)^{3/2}\sigma_x\sigma_y\sigma_z}e^{-\left(x^2/2\sigma_x^2+y^2/2\sigma_y^2+z^2/2\sigma_z^2 \right)}
\label{eq.dens}
\end{equation}
 We can define the optical depth as 
$d = -\ln(P_{\rm out}/P_{\rm in})$ where $P_{\rm in},P_{\rm out}$ are the power incident on and after the atomic cloud. Accounting for the Gaussian transverse distribution of a radially symmetric  laser beam and  diffractive spreading as it propagates along the cloud  we find 
\begin{equation}
d=\frac{4\sigma}{w_0^2}\int_{-L/2}^{L/2} dz\,   \int_0^\infty d\rho\, \rho  n_a(\rho,z)  \frac{1}{1+z^2/z_R^2}e^{-2\rho^2/[w_0^2(1+z^2/z_R^2)]},  
\end{equation}
where $\rho=\sqrt{x^2+y^2},$ $w_0$ is the beam waist and $z_R= \pi w_0^2/\lambda.$
This expression is valid provided  the beam intensity is everywhere small so that there is no saturation of the absorption. When the atomic density has a Gaussian profile as in Eq. (\ref{eq.dens}) we can extend the limits of the $z$ integration to $\pm \infty$ and get the analytical expression
\begin{equation}
\label{eqn.od}
d = N \frac{3\lambda}{\sqrt{2\pi}\sigma_z}\frac{e^{\frac{z_R^2(1+4 \sigma_x^2/w_0^2)}{2\sigma_z^2}}}{\sqrt{1+4 \sigma_x^2/w_0^2}}
{\rm Erfc}\left[ \frac{z_R\sqrt{1+4 \sigma_x^2/w_0^2}}{\sqrt{2}\sigma_z } \right],
\end{equation}
where ${\rm Erfc}(z)=1-(2/\sqrt\pi)\int_0^zdt\, e^{-t^2}$ is the complementary error function. 

\begin{figure}[!t]
\begin{center}
\includegraphics[width=11cm]{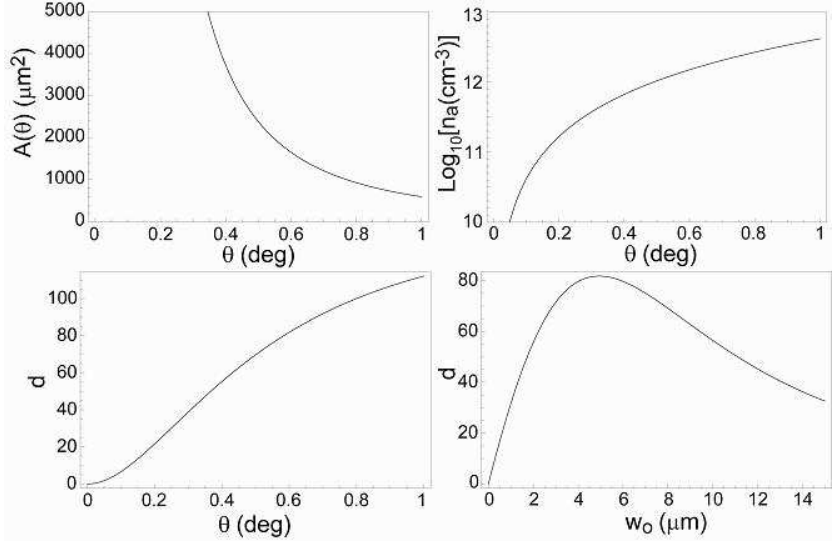}
\caption{Calculated parameters of a compressed pencil shaped cloud: a) Lattice well area, b) peak density, and c) optical depth  for a probe beam with $w_0=5~ \mu \rm{ m}$ as a function of the lattice beam crossing angle $\theta.$ The calculations assumed $w_z=0.18~\rm cm,$ a relative temperature of $k_B T_a/U_0=0.05$ and that 80\% of the $N=5 \times 10^4$ atoms initially in the MOT are captured into the lattice site.  Frame d) gives the optical depth as a function of the probe beam waist $w_0$ for $\theta=0.6~\rm  deg$.  }
\label{fig.od}
\end{center}
\end{figure}

Figure \ref{fig.od} shows the calculated well area, peak density, and optical depth as a function of lattice beam angle $\theta$ and probe beam waist $w_0$ for $N=5 \times 10^4$ atoms initially in the MOT. 
 At $\theta=0.6~\rm deg.$ the peak density reaches the photon rescattering  limit of about  $1.5 \times 10^{12}~\rm cm^{-3}$, but the optical depth along $z$ is $d=82.5.$  This is a factor of about 1800  increase compared to the value in the initial MOT, and is large enough to allow close to complete absorption of a photon in the  atomic cloud. The radial optical depth is less than unity   which implies that the cloud is optically thin for radially directed photons so heating due to photon rescattering will be minimized\cite{Prenti2004}.   This approach to compression is an alternative to anisotropic magnetic 
traps\cite{Schmiedmayerpencil}.

\section{Summary}
\label{summary}

In conclusion we have analyzed protocols for compressing cold atom clouds using time-dependent optical lattices. The combination of a dynamically changing lattice and optical molasses can effectively create highly localized samples with large optical depth.   While this paper focuses on two-dimensional compression, three-dimensional compression is also possible by adding two more traveling waves whose bisector is along $\hat z$.  The result is a three-dimensional lattice which compresses in the same manner as the two-dimensional cases studied here. 
Our numerical  simulations  show that  a high percentage of atoms can be compressed from an atomic cloud with a standard deviation of $\sigma \sim 250 ~\mu\rm m$ to a lattice site with area $A\sim 35\times 35 ~\lambda^2$.  This results in  a pencil shaped optical trap with optical depth $d>80.$ An atomic sample confined in this way is well suited for experiments requiring strong interactions   between light and atomic ensembles.  Further compression of a small number of atoms is also possible down to an area $A \sim \lambda^2$, provided that the density limits set by photon scattering and collisions are not exceeded.

This work was supported by an Advanced Opportunity Fellowship from the University of Wisconsin, NSF grant 
CCF-0523666, and the ARO-DTO  .

\end{document}